\documentstyle[11pt,aaspp4,Flushrt]{article}

\begin{document}
\title{HIGH OBSERVED BRIGHTNESSES IN RADIO JETS}
\author{M. SPADA\altaffilmark{1} M. SALVATI\altaffilmark{2}
AND F. PACINI\altaffilmark{1,2}}
 
\altaffiltext{1}{Dipartimento di Astronomia e Scienza dello Spazio,
		 Universit\`a di Firenze, L. E. Fermi~5, I-50125
		 Firenze, Italy}
\altaffiltext{2}{Osservatorio Astrofisico di Arcetri,
		 L. E. Fermi 5, I-50125 Firenze, Italy}

\begin{abstract}

We study the variability properties at radio frequencies of the jets
thought to be typical of Active Galactic Nuclei, i.e. bulk Lorentz factor
$\Gamma\approx$10, and incoherent synchrotron emission. We assume that
the radiating electrons are accelerated at shocks within the jet, and that
these shocks have a suitable (conical) geometry. In our framework, we can
reproduce the variability pattern of Intra Day Variables (20\% in less
than 1 day) as long as the observed brightness temperature is $T_{B} <
3~10^{17}$~K. The only stringent condition is an injection timescale
of the perturbation shorter than the variability timescale; the geometric
condition on the viewing direction is not especially severe, and agrees with
the observed occurrence of the phenomenon. For higher values of $T_{B}$
coherent processes are probably necessary.

\end{abstract}

\section{INTRODUCTION}

One of the defining features of Active Galactic Nuclei (AGN's) is flux 
variability over most of the electro\-magnetic spectrum, with observed 
timescales which range from minutes up to several years (\cite{gg});
the information content of the
variations makes them a fundamental clue for the understanding of AGN's.
In a previous paper (\cite{aa}, hereafter Paper~I) 
we have considered the extremely fast variations (duration of about 20
minutes) observed in the TeV range from Mkn~421 (\cite{bb}).
These flares imply an extremely compact emitting region and a very 
high photon-photon opacity, unless the source is moving with bulk 
Lorentz factor $\Gamma\approx 100$ (see also \cite{mm}).
We have shown that extreme jet parameters
are not needed, however, if the photon distribution is anisotropic
in the comoving frame; such a distribution is a natural consequence of
conical shocks, with opening angle $\approx 1/\Gamma$. 

At the opposite end of the e.m.
spectrum, similar problems are presented by fast radio variations.
Although radio emission occurs at much larger scales than TeV emission
($10^{19}-10^{20}$~cm versus $10^{16}-10^{17}$~cm), variability up to a 
level $\approx$ 20\% has been observed in various sources with timescales 
shorter than a day, the so called Intra Day Variability (IDV) (\cite{cc},
hereafter WW). The observations imply brightness temperatures up to
$10^{18}-10^{19}$ K, well above the Compton limit $10^{12}$~K
(\cite{pp}).
Because of the relativistic motion of the source the intrinsic $T^c_B$ 
is $\Gamma^3$ lower than the brightness temperature derived from the 
light curves (WW; \cite{nn}).
Again, bulk Lorentz factors $\Gamma\approx 100$ would be
required to cure the problem (see also \cite{zz}).
VLBI observations indicate $\Gamma$ $\approx$ 10 only (\cite{vv};
\cite{ii}), so the intrinsic $T^c_B$ remains well above $10^{12}$~K, 
even after the beaming effects have been corrected for.

Our purpose in this paper is to show that the model developed 
earlier to interpret the high energy variability may be relevant 
to the radio variability as well, with some appropriate modifications.
We shall show that in this way one can account for $T_B$ up to several times
$10^{17}$ K. Higher values of $T_B$ wich have been claimed would
however defy our interpretation. This is especially true for PKS0405--385
(\cite{dd}), which varied on timescales of minutes
with $T_B\sim10^{21}$ K. Barring a relativistic jet with $\Gamma
\sim10^3$, here one is compelled to accept intrinsically high brightnesses,
due perhaps to coherent processes.

\section{DESCRIPTION OF THE MODEL}

Light travel time effects are important in the modelling of variability
since they change the observed time with respect to the coordinate time
by the factor $1-\beta{\rm cos}\theta$, with $\beta$ the velocity of the
source in units of $c$, and $\theta$ the angle between the velocity vector
and the line of sight. If the relevant source velocity is a physical velocity,
one recovers the well known ``shortening'' factor $\sim 1/\Gamma^2$. 
In our scheme, the perturbation is modelled as a slab of extra electrons
superposed on the steady jet, and flowing down the jet with the same bulk 
Lorentz factor $\Gamma$; the extra electrons radiate only after having passed 
through an oblique (conical) shock, where they are accelerated to relativistic
energies in the comoving frame; that is, instead of radiating simultaneously,
they radiate in succession according to their distance from the shock apex
(see Fig.~1 for details). The relevant source velocity, then, is the
phase velocity of the intersection point between the slab and the shock;
calling $\alpha$ the incidence angle, we rewrite the 
``shortening'' factor as $1-\beta{\rm cos}(\theta-\alpha)/{\rm cos}\alpha$,
with $\beta$ the physical velocity corresponding to $\Gamma$, and $\beta/
{\rm cos}\alpha$ the phase velocity. The viewing angle is measured with 
respect to the phase velocity, i.e. from a generatrix of the cone, and is
$(\theta-\alpha)$ if we measure $\theta$ with respect to the physical
velocity. As already pointed out in 
Paper~I, arbitrarily fast variations can be observed when ${\rm cos}\alpha
=\beta$ and $\theta=\alpha$.

Oblique shocks with $\alpha\approx 1/\Gamma$ are indeed expected in 
relativistic jets. If the external pressure drops suddenly, the jet
decollimates and recollimates through a series of alternating conical shocks
with this kind of aperture (Bowman 1994; G\'omez et al. 1997). Furthermore,
radio jets are observed to bend away from their initial direction 
already at the parsec scale; these deviations are probably amplified by
projection effects, and are due to grazing collisions between the jet and
the irregularities of the ambient medium. In this context, one should note
that the transverse pressure exerted by the jet on the ambient medium
is made up of two parts, the comoving pressure and the ram pressure, of
which only the latter increases strongly with $\Gamma$ for incidence
angles larger than $1/\Gamma$. It is then plausible that the external 
medium will give way until angles of this magnitude are reached. Of course,
the shock will not be conical in this case; however, the variability is
dominated by that part of the perturbation which is closest to the line
of sight. All other parts add up on much longer timescales and contribute
to the ``steady'' emission of the jet, so that a precise treatment of their
timing is unnecessary.

The condition that the viewing angle $(\theta-\alpha)$ be small is not very
restrictive: the sources which we study are selected a priori to belong
to the blazar class, and the line of sight is selected a priori to be 
aligned with the jet at angles $\theta \leq 1/\Gamma$
(Urry \& Padovani 1995); within the ``blazing'' cone,
the major part of the solid angle resides close the cone surface, $\theta
\approx 1/\Gamma$, which is our requirement for fast variations.

The length of the injection time determines the thickness of the slab
representing the perturbation, and constitutes a firm lower limit for the
variability timescale. It is tenable that the injection time is determined
by the central engine, the black hole, and that the flow is modulated
with the hole crossing time even at distances much larger than the hole 
radius. Here we follow an analogy with the models invoked for gamma-ray
bursts (Rees \& M\'esz\'aros 1992, 1994): the hyper\-relativistic blast
wave radiates the burst at a fraction of a light year, with a ``global''
timescale determined by the distance times $1/\Gamma^2$; on a finer scale,
however, the light curve is modulated with a time typical of the central
engine, a neutron star, whose dimensions have remained imprinted in the
flow notwithstanding the enormous expansion. In the case of AGN's, this
minimum modulation time could be of a few minutes, compatible with the
fastest variations observed.

Fast variability at radio frequencies, in comparison with gamma-ray
frequencies, has additional features which must be considered. The 
lifetime of the radio emitting electrons is much longer than the gamma-ray
one: in the former case, then, at any given coordinate time radiation
is emitted not only by the electrons which have just been shocked, but
also by older electrons shocked previously, which (see Fig.~1) travel at
different distances from the jet axis. Given the very long lifetime of
the relevant electrons, the emission region can cover the entire cross
section of the jet, even if one includes the decline of the emissivity
due to the lateral expansion of the flow. The observed light curve is
severely smoothed for a generic line of sight. In the following, we show
that in a region of the parameter space fast variability is nonetheless
possible, and that this region is ample enough to account for the observed
statistics.

A final caveat is related with the duty cycle of fast variability at radio 
frequencies. Intra Day Variables exhibit long periods of continuous activity,
with many outbursts superposed at any given time, while at gamma-ray 
frequencies the outbursts seem isolated (observational constraints might 
play a role here). Our model is better suited to account for
isolated outbursts, since the unfavored parts of each slab emit on longer 
timescales, and superpose with other slabs to raise the ``steady'' level
of the source. The fast varying emission is diluted to a small fraction of
the total when many slabs are seen simultaneously. This is in qualitative
agreement with the data, since the gamma-ray flux varies by large factors,
whereas IDV is of the order of a few percent only.
At any rate, a quantitative analysis
requires a more realistic description than that of Paper~I:
here we include properly the dynamics of the flow across the shock, and
the angular distribution of the radiation after the shock.

\section{THE FORMALISM OF THE MODEL}

The flow takes place within a funnel with rigid walls. The walls have an 
opening angle $\psi$ with respect to the $z$ axis, the symmetry axis of the 
system, and a relativistic beam of cold plasma impinges on them at an 
angle $\eta$ (Fig.~\ref{fig1}).  
Before the shock the bulk velocity is $v_1\approx c$ and the Lorentz factor
$\Gamma_1>>1$.
Since the bulk velocity after the shock, $v_2$, must be parallel to the
wall, the Rankine-Hugoniot conditions (see, e.g., \cite{oo})
establish the Lorentz factor
downstream, $\Gamma_2$, and the position of the shock with respect to the 
initial direction of the beam -- i.e. the angle $\alpha$ of Fig.~\ref{fig1}.   
In the frame $S^*$ where the shock is perpendicular 
the Lorentz factor upstream is $\Gamma_1^*=\sqrt{\Gamma_1^2\sin^2\alpha+
\cos^2\alpha}$, with $\Gamma_1^*\approx \Gamma_1$ for angles
$\alpha$ near $\pi/2$ and $\Gamma_1^*\leq \sqrt{2}$ when $\alpha \leq 
1/\Gamma_1$. The shock is not relativistic in general, so the exact solution
of the Rankine-Hugoniot conditions is complicated, and we approximate 
it as follows:
\begin{equation}
v_2^*=\frac{c}{3}\sqrt{1-\frac{41}{25}\frac{1}{(\Gamma_1^*)^2}+
\frac{16}{25}\frac{1}{(\Gamma_1^*)^4}}
\end{equation}
This expression reproduces the two limiting solutions:
$v_2^*=v_1^*/3\approx c/3$ for a relativistic shock
($\Gamma_1^*>>1$); and $v_2^*=v_1^*/5$ in the opposite case
($\Gamma_1^*\leq \sqrt{2}$), under the assumption that the electrons
become relativistically hot even if the protons do not.
In the laboratory frame, finally, the Lorentz factor downstream is:
\begin{equation}
\Gamma_2=\frac{1}{\sqrt{1-\left(\frac{v_1}{c}\right)^2\cos^2\alpha
+\left(\frac{v_2^*}{c}\right)^2
(1-\left(\frac{v_1}{c}\right)^2\cos^2\alpha)}}
\end{equation}
where $\alpha$ is determined by making $v_2$ parallel to the wall:
\begin{equation}
\tan{\epsilon}=\frac{v_2^*}{v_1\cos{\alpha}}
\sqrt{(1-\left(\frac{v_1}{c}\right)^2\cos^2\alpha)}
\end{equation}
and $\epsilon=\alpha-\eta$ is the angle between the shock and the wall.
\placefigure{fig1}

After the shock the electrons radiate effectively for a distance $\Delta s
\leq z$. If $t$ is the coordinate time, the plasma meets the shock in
$z=v_1t\cos(\psi-\epsilon)/\cos\alpha$, and the emission region is a
ring in the $x-y$ plane with internal radius
$r=v_1~t~\sin(\psi-\epsilon)/\cos\alpha$. The non--zero particle lifetime 
gives the width of the ring : 
$\Delta r=\Delta s~(\sin\psi-\cos\psi~\tan(\psi-\epsilon))$ 
(Fig.~\ref{fig1}).  The relation
between the coordinate time $t$ and the observed time $t_o$ ( the time at
which a given wavefront reaches the observer) yields the
infinitesimal surface element of the ring in $t$ which contributes 
to the radiation observed in $t_o$:
\begin{equation}
\Delta A(t_o,t)=\Delta y(t_o,t)~|\frac{dx}{dt_o}|_t
\label{1}
\end{equation}
\begin{equation}
\Delta y(t_o,t)=\sqrt{(r+\Delta r)^2-x^2}-\sqrt{r^2-x^2},~~~
x=-\frac{c}{\sin\theta}[t_o-t(1-\frac{v_1}{c}
\frac{\cos(\psi-\epsilon)} {\cos\alpha}\cos\theta)]
\end{equation}
The integral of Eq. \ref{1} with respect to $t$ -- weighted with the
radiation pattern $P$ of Eq.~\ref{diagr2} -- gives the area $A_w(t_o)$ of 
the source region that generates the flux in $t_o$, $F(t_o)$; this region is 
still ring--shaped, but lies in a plane inclined at an angle $\xi$
with respect to $z$,
$\tan\xi=[(\cos\alpha/\cos(\psi -\epsilon)- \frac{v_1}{c}
\cos\theta)/(\frac{v_1}{c}\sin\theta)]$.
The flux in $t_o$ is proportional to $A_w(t_o)$, and we can   
describe the light curves at different frequencies by an appropriate 
choice of $\Delta s$.  

The high frequencies are radiated by energetic electrons, which cool
very quickly so that $\Delta r << r$; then the model predicts sharp flares 
with timescales $<< \Delta t_z$ for $\alpha\approx 1/\Gamma_1$ and for lines
of sight close to a generatrix of the shock, $\theta\approx\theta_p=\psi
-\epsilon$ (Paper~I). At low frequencies the lifetime of the particles increases 
and $\Delta r$ increases too, until it becomes comparable to $r$. 
The light curve is smoothed, and the shortest variability timescale 
approaches the crossing time $\Delta t_z$. 
In order to observe sharp temporal features
the viewing angle $\theta$ must be close to $\theta_p$, and at the same
time be such that $\xi(\theta)$ is close to $\psi$. In this geometry 
$\vec{\Delta s}$ is parallel to the surface $A(t_o)$ that contributes to 
the radiation at a given $t_o$: the photons emitted by the electrons 
during their lifetime reach the observer at the same instant, and the 
light curve is independent of $\Delta s$.
The angle $\theta$ for which this occurs is:
\begin{equation}
\cos{\theta_{\xi}}=\frac{\cos{\alpha}\pm|\tan{\psi}|
\sqrt{\left(\frac{v_1}{c}\right)^2\cos^2{(\psi-\epsilon)}
(\tan^2{\psi}+1)-\cos^2{\alpha}}}
{\frac{v_1}{c}\cos{(\psi-\epsilon)} (1+\tan^2{\psi})}
\end{equation}
Fig.~\ref{fig2} shows the theoretical light curves 
for different viewing angles
around $\theta_{\xi}$. If we assume $\Gamma_1=15$, $\alpha=1/\Gamma_1$ and
$\psi=0.1$, then $\theta_{\xi}\approx 0.04$, $\theta_p\approx 0.08$,
and $\Gamma_2\approx 10$.
If $r_0$ is the initial jet radius, the light curves are computed
assuming that $\Delta s=10~r_0$:
over this distance the jet radius doubles, and the emissivity drops
by a factor of several. The variability is about 25\% on a time
scale $\approx 0.1\Delta t_z$, if $\theta$ is between 0.04 and 0.06.
\placefigure{fig2}

The light curves in Fig.~\ref{fig2} are computed with a realistic
angular distribution of the radiation. In the frame comoving with
the plasma the magnetic field upstream is assumed isotropic,
downstream the magnetic field component parallel to the shock $B_{//}$
dominates the normal one $B_{\perp}$, even if $\Gamma_1^*\sim 1$.
Including only $B_{//}$ and transforming back to the laboratory frame, we get
\begin{equation}
\frac{dP(\chi,\phi)}{d\Omega}\propto
\left[1+\left(\frac{\sin{\epsilon_c}(\cos{\chi}-\frac{v_2}{c})}
{1-\frac{v_2}{c}\cos{\chi}} + \frac{\cos{\epsilon_c}
\sin{\chi}}{\Gamma_2(1-\frac{v_2}{c}\cos{\chi})\sqrt{1+\tan^2{\phi}}}
\right)^2\right] \frac{\delta^3}{\Gamma_2}
\label{diagr2}
\end{equation}
where $P$ is the power emitted in a unit solid angle in the direction 
defined by $\chi$ and $\phi$, $\delta=[\Gamma_2~(1- v_2/c~\cos{\chi})]$ 
is the Doppler factor, $\epsilon_c=\tan(\alpha-\eta)/\Gamma_2$ is the angle 
between the shock and the wall in the comoving frame after the shock, and
\begin{equation}
\cos{\chi}=\cos{\theta}\cos{\psi}(1+\frac{x}{r}
\tan{\theta}\tan{\psi})
\end{equation}
\begin{equation}
\tan{\phi}=\frac{\tan{\theta}\sqrt{1-(x/r)^2}}{\sin{\psi}}
\end{equation}
If $x'$,$y'$,$z'$ are cartesian coordinates with $z'$ along $v_2$
(parallel to the wall), the $z'x'$, and $x'y'$ sections of the radiation
pattern for $\Gamma_1$=15 and $\alpha=1/\Gamma_1$ are shown in
Fig.~\ref{fig3}. If the opposite walls of the funnel diverge by
an angle $\geq$ the width of the radiation pattern, the long--lasting tail
of a flare due to the `unfavored' side of the funnel is much suppressed;
this is important when modeling light curves which, as the radio ones,
show repeated flares with a high duty cycle.
\placefigure{fig3}

\section{COMPARISON WITH THE DATA AND DISCUSSION}

Fig.~\ref{fig4} shows the comparison of a theoretical light curve with the
observations of S5~0716+714 (\cite{ee}).
This source is classified as a BL--Lac object, with a bright
(S$_{5~GHz}>1$ Jy) and flat radio emission ($S_{\nu}\propto \nu^{\alpha}$
with $\alpha=0.42$ between 1.5 and 5 GHz, \cite{ff}), and
a significant optical polarization.
The optical spectrum is completely featureless and the redshift 
determination is difficult.  The absence of any host galaxy rules out
redshifts $z< 0.25$ and it is generally assumed $z>0.3$.
\placefigure{fig4}

The light curve in Fig.~\ref{fig4} represents the VLA 
observations of February 1990. 
The source was observed at $\lambda=6$ cm every two hours,
with a regular sampling (except for weekly maintenance gaps).    
Normalized to the average brightness, the light curve shows intraday
variability, in particular in the first  part of the campaign.

The theoretical light curve in the lower panel was obtained by
convolving many curves like those of Fig.~\ref{fig2}, 
with random initial times
and amplitudes. The time separation between successive micro\-flares 
has a flat distribution between 0.025 and 0.075 $\times \Delta t_z$,
the amplitude has a flat top distribution of relative width 20\%.
The parameters of the elementary curves are equal to the those 
of Fig.~\ref{fig2}, with $\theta=0.04$.
The choice of the intrinsic time $r_0/(c\Gamma_1)$ sets the physical
size of the source and the variability timescale; here we have taken
$r_0/(c\Gamma_1)$=13 days.
Since similar results are obtained with larger
$\theta$'s up to 0.06, the visibility fraction is
$(0.06^2-0.04^2)\times \Gamma_2^2\approx$0.24, in good agreement  
with the occurrence of Intra Day Variability in the Effelsberg catalog.
Indeed, our results are valid when the line of sight is at a small
angle with respect to the shock surface; since the
line of sight must be a priori within $1/\Gamma_2$ of the same surface,
in order that the source shows blazar features, the visibility fraction is
substantial. The model by \cite{ll} is similar to ours in requiring an 
injection timescale shorter than the variability timescale; on top of this,
the line of sight must be at a small angle with respect to the symmetry 
axis, instead of the shock surface, and the visibility fraction is 
diminished with respect to ours by a factor of order $1/\Gamma_2$.

\begin{figure}\vspace*{-4cm}
\centerline{\psfig{figure=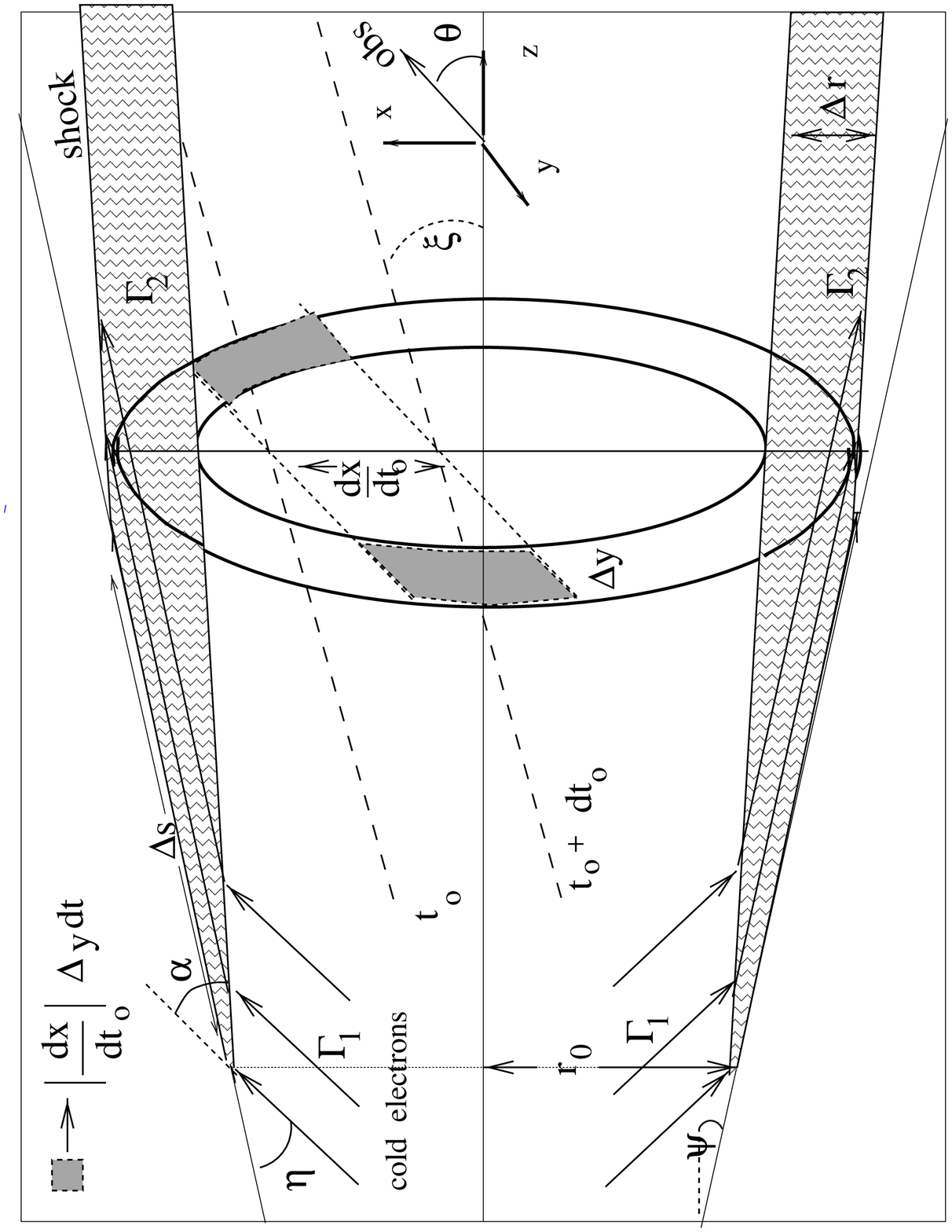,width=15cm,angle=-90}}
\caption{The source geometry. A beam of cold electrons ($\Gamma_1$)
impinges on the jet walls at an angle $\eta$ . A shock is formed 
at an angle $\alpha$ with the initial velocity of the particles, 
and the electrons became relativistic in the comoving frame (dashed
part). They move along the jet walls ($\Gamma_2$) and radiate for a
distance $\Delta s$. The viewing angle is $\theta$.
The ring in the figure is the emission region at the coordinate
time $t$. Only the filled part of this ring 
($\Delta A=\Delta y |dx/dt_{o}|_t$) contributes to the flux $F(t_{o})$.
The two dashed lines indicate the positions
of the planes (normal to the page, and inclined by $\xi$ 
with respect to $z$), from which the photons emitted at different
$t$'s arrive together to the observer between $t_{o}$ and 
$t_{o}+dt_{o}$.}
\label{fig1} 
\end{figure}

\begin{figure}\vspace*{-4cm}
\centerline{\psfig{figure=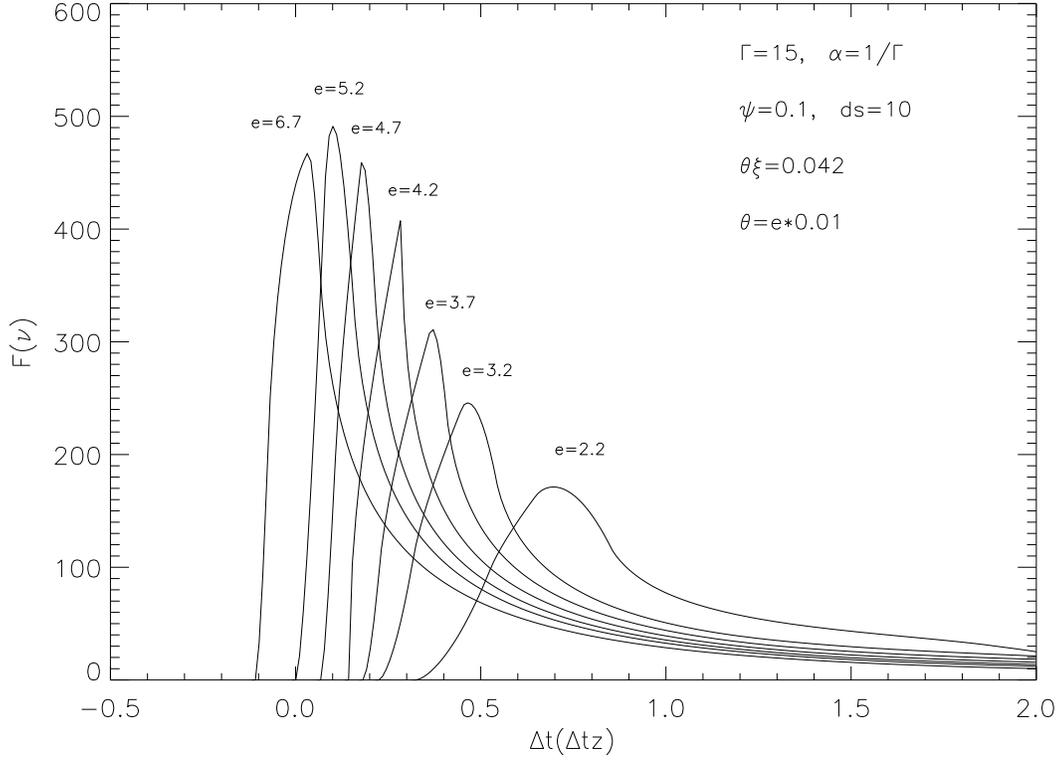,width=15cm,angle=90}}
\caption{The theoretical light curves for different viewing angles around
$\theta_{\xi}=0.042$, with $\Gamma_1=15$, $\alpha=1/\Gamma_1$,
$\psi=0.1$ and $\Delta s = 10~r_0$. The time is normalized to the
crossing time $\Delta t_z$ and the flux $F(t_o)$ to its average
value. If $\theta$ is between 0.04 and 0.06, the variability is about 
$25\%$ on a timescale of $\approx 0.1 \Delta t_z$.}
\label{fig2}
\end{figure}

\begin{figure}\vspace*{-3cm}
\centerline{\psfig{figure=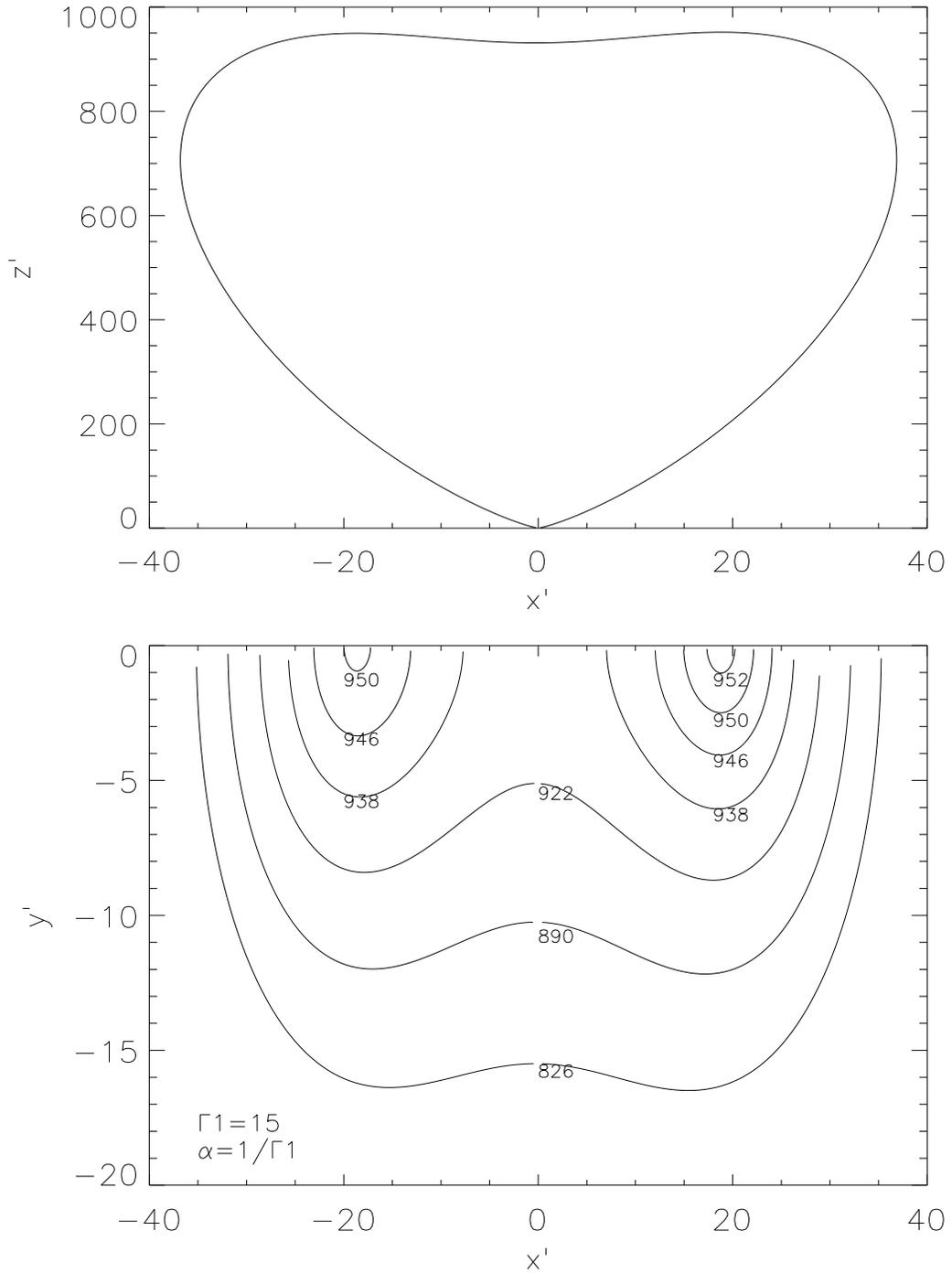,width=15cm}}
\caption{The radiation pattern for $\Gamma_1=15$ and $\alpha=1/\Gamma_1$.
If $x'$,$y'$,$z'$ are cartesian coordinates with $z'$ along $v_2$,  
the upper panel shows the section $x'z'$, and the bottom one the section
$x'y'$.}
\label{fig3}  
\end{figure}

\begin{figure}\vspace*{-3cm}
\centerline{\psfig{figure=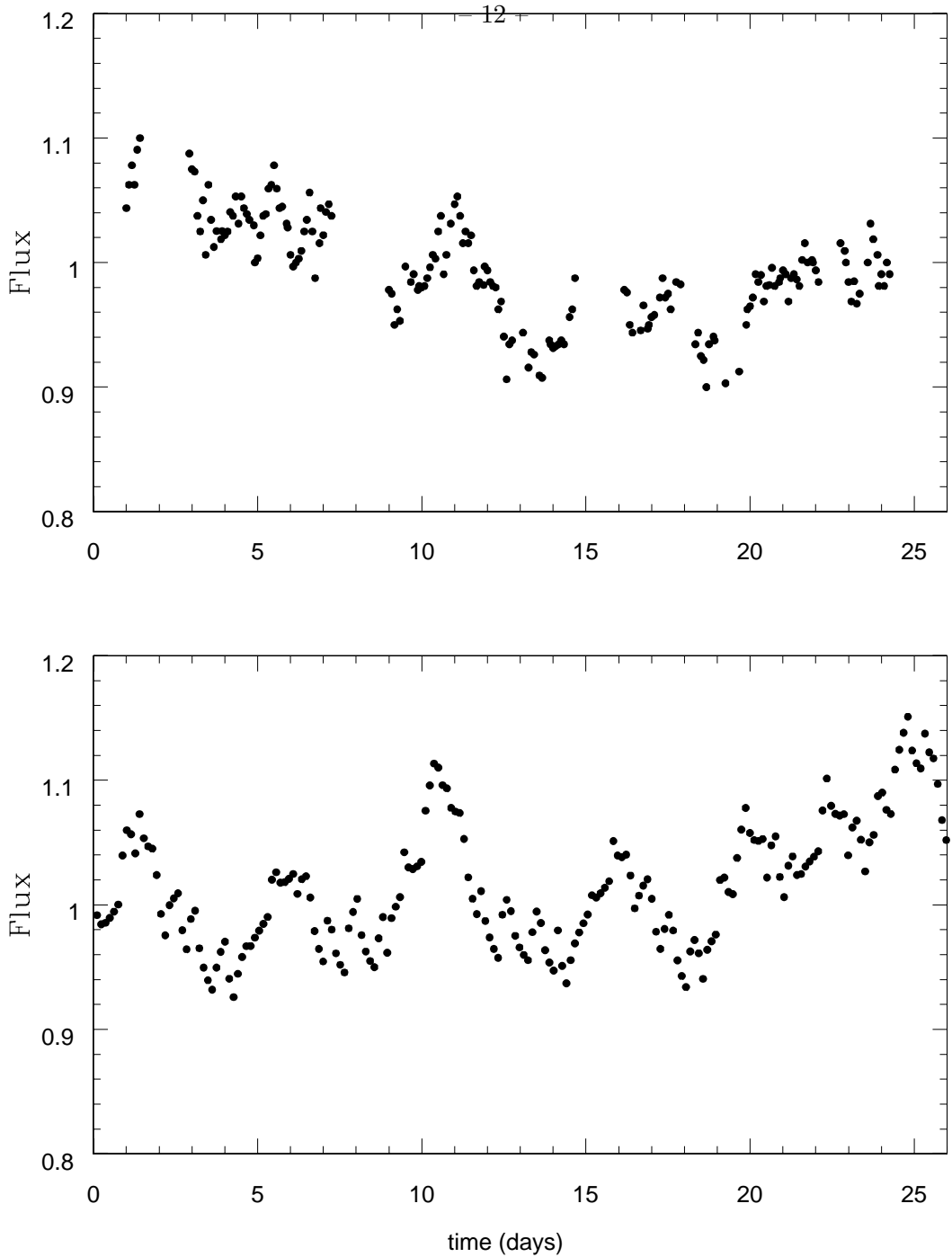,width=15cm}}
\caption
{The observed (top panel) and theoretical (bottom panel) light curve 
of S5~0716+714, at 5 GHZ. The data are taken from Wagner et al. (1996).
The abscissa is the time in days, and the ordinate 
the flux, normalized to the average one. The parameters of the
theoretical curve are: $\Gamma_1=15$, $\alpha = 1/\Gamma_1$, $\Delta s =
10~r_0$, $\psi=0.1$, $r_0/c\Gamma_1=$13 days.}
\label{fig4}
\end{figure}

A final parameter to be chosen is the average flux, to which the curves
are normalized. If the comoving brightness temperature is $T^c_B$, 
$q_0=1/2$ and $H_0=70 \rm ~km ~s^{-1}~Mpc^{-1}$,
the flux emerging from the base of the cone in Fig.~\ref{fig1} can 
be written as
\begin{equation}
F_{\nu}=\frac{2\nu^2(1+z)^3k_B\Gamma_2T^c_B 3\pi[c\Gamma_1~
13~{\rm days}/(1+z)]^2}{(9~{\rm Gpc}\times(1+z-\sqrt{1+z}))^2}
\label{flux}
\end{equation}
The maximum value for $T^c_B$ is 10$^{12}$~K, of course. 
At relatively low redshifts, appropriate to objects like S5~0716+714,
we obtain fluxes of the order of those observed in the S5 catalog, i.e. 
about a Jansky. 

The particular source in question, S5~0716+714, was
chosen as a paradigm of IDV at low redshifts because of the extensive
data base, and from this point of view the performance of the model
is comforting. However, S5~0716+714 is also the only firm example of
correlated optical and radio IDV (WW). Most models
have the high frequency emission coming from the inner jet, at distances
of about 10$^{16}$--10$^{17}$~cm, and the radio emission from further out,
at distances of about 10$^{19}$--10$^{20}$~cm. If the correlation is
accounted for by making cospatial the optical and the radio, then the 
canonical scheme must be abandoned: placing the radio source in the inner
jet would imply extreme brightness temperatures and coherent processes
(see below). Moving the high frequency source further out would be
compatible with the scheme proposed here, but would be difficult to reconcile
with models involving inverse Compton, because of the low density 
of targets; one could perhaps resort to models involving proton induced
cascades (Mannheim 1993).

At redshifts of order 1 and higher, appropriate to quasar IDV's, the fluxes 
predicted by Eq. \ref{flux} are too low. We can obtain reasonable fluxes by
taking higher values for $r_0/(c\Gamma_1)$, at the expense of  
longer variability timescales. These limitations amount to an upper
bound on the apparent brightness temperature of several times 
10$^{17}$~K (with the same definitions of WW). We conclude that,
while a non--negligible fraction of IDV's can in principle be accounted
for by our model, the extreme events, and especially the 10$^{21}$~K 
flare observed in PKS0405--385, may involved alternative scenarios as
coherence.

\acknowledgements
This work was partly supported by the Italian Space Agency
(ASI) throughout grants ARS--96--66 and ARS--98--116/22.

\newpage

\input PSFIG

\newpage

\input PSFIG
\newpage

\newpage

\end{document}